\documentclass[conference]{IEEEtran}

\usepackage{graphicx}

\begin{document}
\title{On Double-Sided QR-Codes}

\author{\IEEEauthorblockN{Alexey Tikhonov}
\IEEEauthorblockA{Yandex Technology GmbH, Germany}
altsoph@gmail.com
} 

\maketitle

\begin{abstract}
    Due to the widespread adoption of the smart mobile devices, QR codes have become one of the most-known types of 2D codes around the world.
    However, the data capacity properties of modern QR codes are still not perfect. To address this issue, in this paper, we propose a novel approach to make \textit{double-sided} QR codes, which could carry two different messages in a straight and mirrored position. To facilitate the process of creation of such codes we propose two methods of their construction: the brute-force method and the analytic solution.
\end{abstract}

\IEEEpeerreviewmaketitle

\section{Introduction}
Originally developed for automotive industry tasks in the early 1990s, QR Codes or two-dimensional barcodes are used to encode and decode data at a rapid rate. The speed of scanning, the powerful error correction and the readability from any direction gave QR codes \cite{CITE1} \cite{CITE2} great popularity in common life. Using camera phones and appropriate applications to read 2D barcodes for various purposes is currently a widely used approach in practical applications \cite{CITE3}. Anyone with a camera phone equipped with the correct reader application can scan the image of the QR code to display text, contact information, connect to a wireless network, or open a web page in the telephone's browser.

However, the data capacity of modern QR codes is still very limited, which hinders possible extensions of their applicability, e.g. adding authentication mechanisms for protection from information leakage \cite{CITE4}. There are several approaches which try to address this capacity problem, e.g. usage of multicolored high-capacity QR codes \cite{CITE5} or IQR codes with increased density. Still, such approaches imply making at least some changes in the scanning software, which is very difficult taking into account a number of different scanning applications in existence.

\begin{figure}[!htb]
	\includegraphics[width=0.49\textwidth]{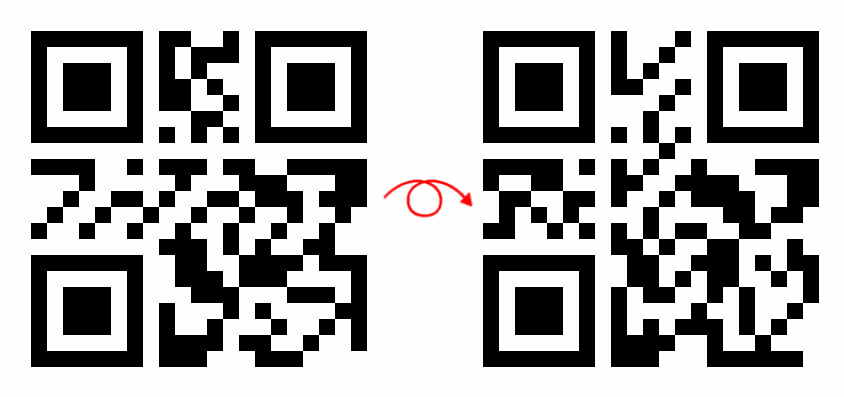}
	\caption{Code with 'HARRY' message, whose mirror version reads as 'BOVIK'}
	\label{fig:QRBOVIK}
\end{figure}

Instead of changing the scanning software, we propose the usage of specially crafted QR codes, which could carry two different messages in a straight and mirrored position (see Fig.\ref{fig:QRBOVIK} for an example). To facilitate the process of creation of such codes we propose two methods of their construction: the brute-force method and the analytic solution.

\section{Anatomy of QR Codes}
\label{sec:ANAT}

Let us refresh some basic information about the QR code structure. 
Here and further we will consider the simplest type of QR codes, Version 1-L. This means the code will have only 21x21 pixels and the lowest possible error correction level. Each such code consists of several different areas. Namely, there are some fixed pixels, the special control code area, the data area, and the error correction zone.

\begin{figure}[!htb]
	\includegraphics[width=0.49\textwidth]{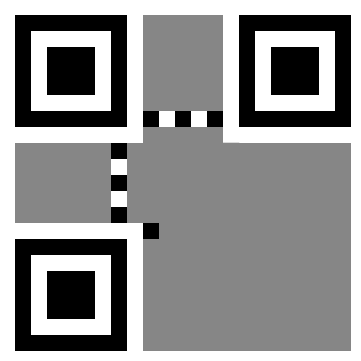}
	\caption{Fixed pixels of a QR code, Version 1-L}
	\label{fig:QR1}
\end{figure}

\paragraph{Fixed Pixels} A part of the code is always fixed and filled with so-called function patterns. They are used for the code localization: the reader algorithm bases on these patterns to understand the position and the orientation of code. Check Fig.\ref{fig:QR1} for the positions of fixed pixels.

\begin{figure}[!htb]
	\includegraphics[width=0.49\textwidth]{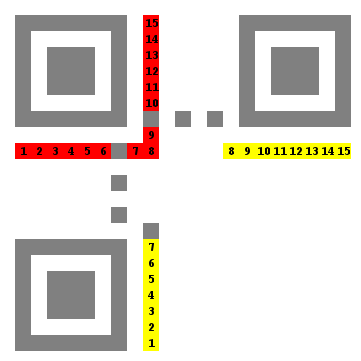}
	\caption{Location of control code bits of a QR code, Version 1-L}
	\label{fig:QR2}
\end{figure}

\paragraph{Control Code Area} The most important variable part of a QR-code is so-called control code. It contains only 5 bits of control information which specify parameters of a further decoding process. It’s vital, so it’s highly protected: there is 3-fold redundancy for the Bose-–Chaudhuri-–Hocquenghem code correction and the whole thing repeats on the code twice in two different places (Fig.\ref{fig:QR2}).

\begin{figure}[!htb]
	\includegraphics[width=0.49\textwidth]{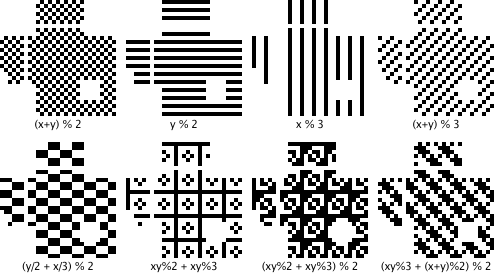}
	\caption{8 possible XOR-masks}
	\label{fig:QR4}
\end{figure}

Among these 5 bits of the control code, two of them encode the error correction type (in our case it will be “01” for L-type which stands for “Low”). Another three bits contain the code of one of 8 possible XOR-masks applied to the main payload and error correction areas (possible masks are shown on Fig.\ref{fig:QR4}). The last 10 bits added for the BCH(15,5) error correction.

\begin{figure}[!htb]
	\includegraphics[width=0.49\textwidth]{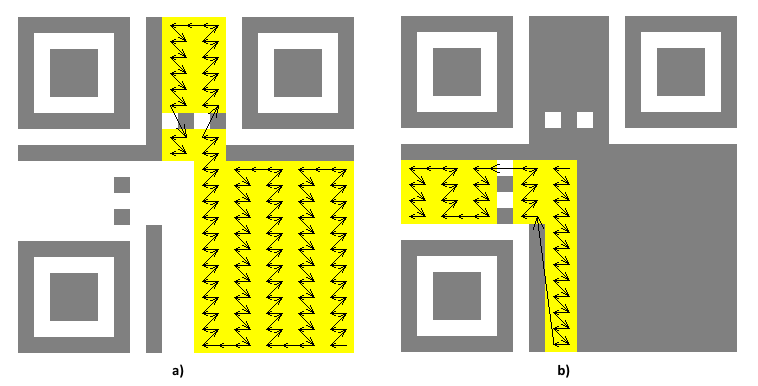}
	\caption{a) payload data bits, b) error correction bits of QR code, Version 1-L}
	\label{fig:QR3}
\end{figure}

\paragraph{Payload Data Area and Error Correction Zone} The rest of the space is divided between the payload data and error correction data. In case of QR code, Version 1-L we have 152 bits of actual data (Fig.\ref{fig:QR3},a)) and 56 bits of the error correction data occupy the rest (Fig.\ref{fig:QR3},b)).

\paragraph{Payload Structure} The payload data itself has some internal structure which depends on the used encoding mode. The QR-code standard gives us a choice from several different encoding modes:
\begin{itemize}
  \item Numeric Mode -- only numbers,
  \item Alphanumeric Mode -- 2 symbols in 11 bits, no cases, short alphabet,
  \item Byte Mode -- typical 8 bits per char,
  \item Kanji Mode -- 16 bits per char, wide alphabet,
  \item ECI Mode and others -- too complex for our purposes.
\end{itemize}

Let's say we use the Alphanumeric Mode because it’s thrifty and has most of the useful chars. How a typical payload data will look like?
\begin{itemize}
  \item Code of encoding mode, 4 bits: 0010 for Alphanumeric Mode,
  \item Length of data in characters, for 1-L Code with Alphanumeric encoding this field has 9 bits length,
  \item Data itself, 2 symbols in 11 bits, 6 bits for last odd symbol,
  \item Terminator. The terminator itself has a complex structure:\begin{itemize}
  \item Terminating sequence “0000” (4 bits),
  \item Additional zeros for 8-bit padding (0 -- 7 bits),
  \item Filling pattern “11101100 00010001” till the end of data (whole 19 bytes).
\end{itemize}
\end{itemize}

\paragraph{Error Correction Notes} using 1-L Version QR code we have the error correction up to 24 bits, but they should be located in up to 3 padded bytes. That's what Reed--Solomon codes usually used for: we could correct a lot of errors as long as they are localized to a small number of fixed spots.

\section{Flipping The Code}
\label{sec:MIRR}

\subsection{Flipping Service Areas}
To make the both sides of code readable we should be aware where different parts of the code map after the reflection along the main diagonal. The first question is -- is it possible at all to build a double-sided code with the correct control structures.

\paragraph{Flipping Static Area} The static area is almost symmetrical and maps into itself except the one pixel called Dark Module according to the standard specification. This Dark Module maps into the 8th bit of the one copy of the control code, so ideally we should prefer to use control codes with the middle bit equal to 1.

\begin{figure}[!htb]
	\includegraphics[width=0.49\textwidth]{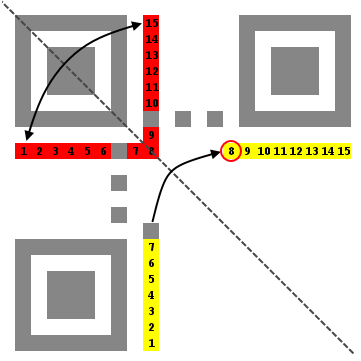}
	\caption{Flipping service areas}
	\label{fig:QR6}
\end{figure}

\paragraph{Flipping Control Code} Except this Dark Module invasion the both copies of the control code map precisely into themselves but \textit{reversed} (Fig.\ref{fig:QR6}).

\par So, ideally, we need the palindromic control code with “1” right in the middle. Also, we want the L-type error correction and we prefer the symmetrical XOR masks (beсause it will be much easier to deal with symmetrical XOR when we will do the mirroring of the payload data) which leaves us with 5 possible XOR masks out of 8.

\subsection{Crafting Handy Control Code}

Is it possible to construct the desired control code value? Yes, if we use the BCH(15,5)’s ability to correct up to 3 bits. And we could do it \textit{on the both sides}. So actually we are looking for a 15-bit binary string which has up to 3 bits difference with the desired code AND up to 3 bits difference with the reversed desired code at the same time.

\begin{figure}[!htb]
	\includegraphics[width=0.49\textwidth]{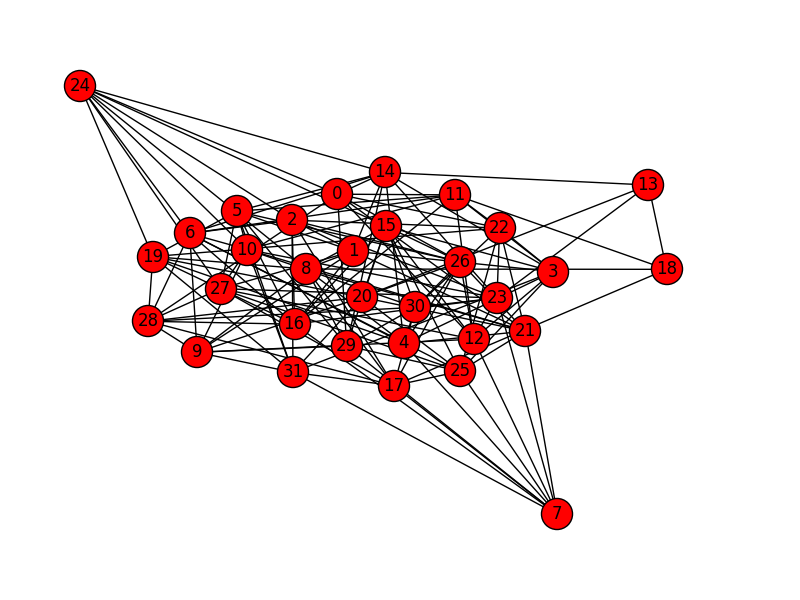}
	\caption{Flip-graph of control codes}
	\label{fig:QR5}
\end{figure}

Since there are only $2^5=32$ different valid codes a brute-force approach could be used to check all vectors inside spheres within a 3-bit radius around each of the valid code. We still will have only $2^5\cdot{15\choose 3}=14560$ possible candidates (actually less, since they are repeating). The results could be presented as an undirected graph (Fig.\ref{fig:QR5}) with 32 correct codes as nodes where the edge between codes A and B exists if there is such a 15 bits string C, so C is within 3-bits radius from A and the reversed(C) is within 3-bits radius from B. 

This graph actually has even the loops, so the best choice for our task would be something like:
$$100101010100001 <=> 100001010101001$$
\par This code:
\begin{itemize}
  \item has only 2 bits difference from its reversed version,
  \item has 1 in the middle bit, which is resistant to Dark Module,
  \item means 1-L error correction and symmetrical XOR mask.
\end{itemize}

\subsection{Injecting Data}

Finally, we need to put our payloads for the both sides into one code and check how heavy they are overlapped. It appears to be a problematic part because the area of the overlap between data areas is 100 bits and covers the beginnings of the both messages. However, things are really better when the payload is short.

\par
\begin{figure}[!htb]
	\includegraphics[width=0.49\textwidth]{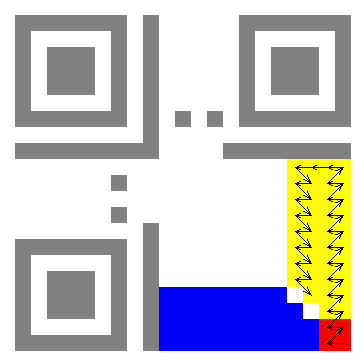}
	\caption{Mapping two short payloads (41 bits each)}
	\label{fig:QR7}
\end{figure}

 Let’s, for example, try to put 5 symbols on each side, i.e. “HELLO” on the first side, and use the Alphanumeric encoding mode:
\begin{itemize}
  \item Encoding mode: 0010
  \item Len: 000000101
  \item Data: 01100001011 01111000110 011000
  \item No terminator: our field tests show nobody cares about the terminator and filler, a value of the Length field is just enough.
\end{itemize}
The total length of our payload is 41 bits. Given the same size for the other payload we will end with the mapping shown in Fig.\ref{fig:QR7}. The whole intersection is only 4 bits and it gets its’ place inside the encoding mode code (“0010” vs “0100”), so the actual difference is only 2 bits so long.

\begin{figure}[!htb]
	\includegraphics[width=0.49\textwidth]{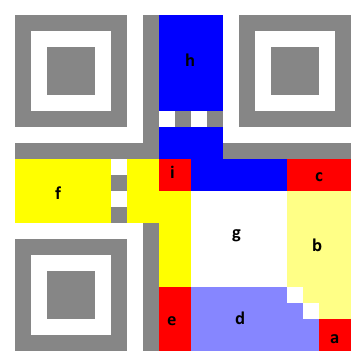}
	\caption{Mapping two short payloads (a+b+c) and (a+d+e) and two error correction zones (f+e+i) and (h+i+c). The conflicts are in the zones a, c, e, i.}
	\label{fig:QR8}
\end{figure}

But we still need to put the error correction data for the both sides as it's obligatory and it couldn’t be changed directly as it’s a complex function of the payload data. Meanwhile, it has big overlaps over itself and over the data area (see Fig.\ref{fig:QR8}). So, for short payloads, we have 2 bits error from the data intersection + 20 new possible overlapped bits from error correction zone. But it does the error correction, so maybe it’s enough to correct itself?

\subsection{Error Correction Knot}
The idea is to use the power of the error correction to correct some problems caused by overlap of the error correction zones. Remember, our code is capable to correct up to 3 padded bytes \textit{for each side}, so if we can put each of our error bits on one or another side and arrange all errors in groups, up to 3 bytes on one side and up to 3 bytes on another, we could make it. 
\par First, let's enumerate important parts of data on the code (see Fig.\ref{fig:QR8} again):

\begin{itemize}
  \item \textbf{a + b + c + d + g + h = data1}
  \item \textbf{a + d + e + b + g + f = data2}
  \item \textbf{f + i + e = control1(data1)}
  \item \textbf{h + i + c = control2(data2)}
  \item \textbf{a, c, e, i} -- the conflicting zones
\end{itemize}

\par The worst problem is what we have $f=control(h,g,...)$ and $h=control(f,g,...)$ at the same time, and $f$ has to match the flipped version of $f$. Since the computation of error correction bits is not so easy to reverse we have no direct control on $f$ or $h$.

\section{Fighting with Error Correction}
\subsection{Bruteforce approach}

Since both $f$ and $h$ depend on $g$ and we are free to change $g$ to anything we want, we could just random search for such value of $g$, which will give us $f$ and $flipped(h)$ similar enough to cover the differences with the error correction.

\subsection{Upper limit}
\begin{figure*}[t]
	\includegraphics[width=0.98\textwidth]{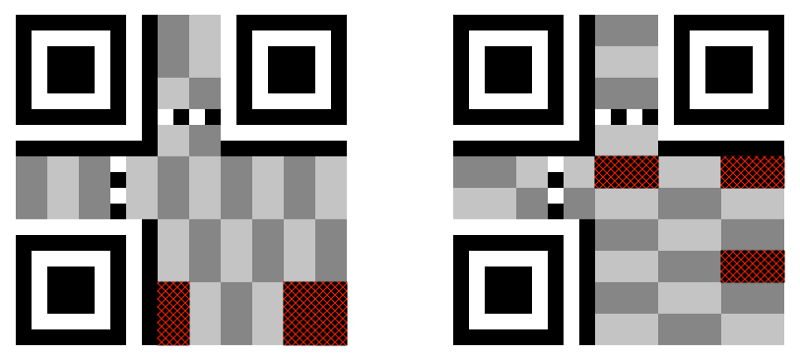}
	\caption{The optimal correction scheme with 3 padded bytes on each side.}
	\label{fig:QR9}
\end{figure*}

Let’s see how many errors we could cover at most with 3 bytes on the each side: Fig.\ref{fig:QR9}.

Such an approach gives us up to 8 symbols message on the one side and up to 11 symbols message on the other one. But since we have almost no freedom degrees left, the bruteforce could take ages. 

\subsection{Analytic Solution}

The QR-code standard uses Reed–Solomon codes for the error correction. Thus the whole error correction area is a known multi-dimensional boolean function of data: we could write it down twice (for the two sides). Then we add more equations which bind the values of the same bit on the different sides.

\par So we have a huge system of linear boolean equations with some free variables in it. Such system can be sold analytically just in milliseconds using, for example, the Gaussian elimination method. Then, setting any values for the free variables we could compute the values for all the bits of our code.

\section{Conclusion}	

In this paper, we propose a novel approach to make double-sided QR codes, which could carry two different messages in a straight and mirrored position. To facilitate the process of creation of such codes we propose two methods of their construction: the brute-force method and the analytic solution. However, we have encountered some technical difficulties, which impose limits on the length of the message. This problem might be addressed in future studies.
    

\bibliography{covert}{}
\bibliographystyle{plain}

\end{document}